\documentclass[sort&compress,final]{aipproc} 

\layoutstyle{6x9}

\newcommand\etal{~et~al.}
\newcommand\arcdeg{\ensuremath{^\circ}}    
\newcommand\kparsec{\ensuremath{\,\mathrm{kpc}}}
\newcommand\keV{\ensuremath{\,\mbox{keV}}}
\newcommand\aj{AJ}          
\newcommand\araa{ARA\&A}    
\newcommand\aap{A\&A}       
\newcommand\apj{ApJ}        
\newcommand\apjl{ApJ}       
\newcommand\mnras{MNRAS}    
\newcommand\nat{Nature}     

\begin{document}


\title{Polarimetry and the High-Energy Emission Mechanisms in Quasar Jets}



\author{M.~Cara}{address={Department of Physics~\& Space Sci., Florida Institute of Technology, Melbourne, FL~32901, USA}}

\author{E.~S.~Perlman}{address={Department of Physics~\& Space Sci., Florida Institute of Technology, Melbourne, FL~32901, USA}}

\author{Y.~Uchiyama}{address={SLAC National Accelerator Laboratory, Menlo Park, CA 94025, USA}}

\author{S.~Jester}{address={Max Planck Institute for Astronomy, 69117 Heidelberg, Germany}}

\author{M.~Georganopoulos}{address={Joint Center for Astrophysics, University of Maryland, Baltimore, MD~21250, USA}}

\author{C.~C.~Cheung}{address={NASA Goddard Space Flight Center, Greenbelt, MD 20771, USA}}
\author{R.~M.~Sambruna}{address={NASA Goddard Space Flight Center, Greenbelt, MD 20771, USA}}

\author{W.~B.~Sparks}{address={Space Telescope Science Institute, Baltimore, MD 21218, USA}}

\author{A.~Martel}{address={Space Telescope Science Institute, Baltimore, MD 21218, USA}}

\author{C.~P.~O'Dea}{address={CIS and Department of Physics, Rochester Institute of Technology, Rochester, NY 14623, USA}}

\author{S.~A.~Baum}{address={CIS and Department of Physics, Rochester Institute of Technology, Rochester, NY 14623, USA}}

\author{D.~Axon}{address={CIS and Department of Physics, Rochester Institute of Technology, Rochester, NY 14623, USA}}

\author{M.~Begelman}{address={Joint Institute for Laboratory Astrophysics, University of Colorado, Boulder, CO~80309, USA}}

\author{D.~M.~Worrall}{address={Department of Physics, University of Bristol, Bristol~BS8~1TL, UK}}

\author{M.~Birkinshaw}{address={Department of Physics, University of Bristol, Bristol~BS8~1TL, UK}}

\author{C.~M.~Urry}{address={Department of Physics, Yale University, New Haven, CT 06520-8121, USA}}

\author{P.~Coppi}{address={Department of Astronomy, Yale University, New Haven, CT 06520-8101, USA}}

\author{\L{}.~Stawarz}{address={KIPAC, Stanford University, Stanford, CA 94305, USA}}

\begin{abstract}
The emission mechanisms in extragalactic jets include synchrotron and various inverse-Compton processes.  At low (radio through infrared) energies, it is widely agreed that synchrotron emission dominates in both low-power (FR~I) and high-power (FR~II and quasar) jets, because of the power-law nature of the spectra observed and high polarizations.  However, at higher energies, the emission mechanism for high-power jets at $\kparsec$ scales is hotly debated.  Two mechanisms have been proposed:  either inverse-Compton of cosmic microwave background photons or synchrotron emission from a second, high-energy population of electrons.  Here we discuss optical polarimetry as a method for diagnosing the mechanism for the high-energy emission in quasar jets, as well as revealing the jet's three-dimensional energetic and magnetic field structure. We then discuss high-energy emission mechanisms for powerful jets in the light of the HST polarimetry of PKS~1136$-$135.
\end{abstract}

\maketitle

\section{Introduction}

The jets of radio-loud AGN carry energy and matter out from the nucleus to cluster-sized lobes, over distances of hundreds of kpc.  While found in only $\sim 10\%$ of AGN, jets can have a power output (including both luminosity and kinetic energy flux) comparable to that of the host galaxy and AGN \cite{Rawlings:91}, and can profoundly influence the evolution of their host galaxy and nearby neighbors. AGN jets are completely ionized flows, and the radiation we see from them is non-thermal in nature. That the radio emission arise from synchrotron radiation is supported by strong linear polarization and power-law spectra seen in both lower- and higher-power jets. However, at higher energies, the nature of the emission from higher-power large-scale jets is less clear.

In low-power FR~I radio galaxies, the optical and X-ray fluxes fit on extrapolations of the radio spectra (e.g., \citep{Perlman:01,Hardcastle:01,Perlman:05}), and high polarizations are seen in the optical (typically $\sim 20-30\%$, \cite{Perlman:99,Perlman:06}) suggesting synchrotron emission.  These jets exhibit a wide variety of polarization properties \cite{Perlman:06,Dulwich:07,Perlman:09}, often correlated with X-ray emission. For example, in the jet of M87 \cite{Perlman:05}, a strong anti-correlation between X-ray emission and optical polarization was found in the knots, accompanied by changes in the magnetic field direction, suggesting a strong link between the jet's dynamical structure and high-energy processes in the jet interior, where shocks compress the magnetic field and accelerate particles in situ. 

For the more powerful FR~II and quasar jets, the nature of both the optical and X-ray emission is under active debate. In many FR~II jets, the optical emission can lie well below an interpolation between the radio and X-rays (e.g., \cite{Sambruna:04}), sometimes by decades (e.g., PKS~0637$-$752, \cite{Schwartz:00,Mehta:09}), resulting in a characteristic double-humped shape of spectral energy distribution. In some jets the optical emission appears linked to the X-ray emission by a common component. This is the case in both 3C~273 and PKS~1136$-$135, where deep HST, Chandra and Spitzer imaging \cite{Jester:01,Jester:06,Jester:07,Uchiyama:06,Uchiyama:07} has shown that a second component, distinct from the lower-energy synchrotron emission, arises in the near-IR/optical and dominates the jet emission at optical and higher energies, at least up to 10 keV. Competing mechanisms have been proposed: either synchrotron radiation from very high-energy particles or inverse-Comptonization (see \cite{Kataoka:05,Harris:06}), however, the nature of this component cannot be constrained by multi-waveband spectra alone \cite{Georg:06}.

\section{Polarimetry as a Diagnostic Tool}

Polarimetry is a powerful diagnostic for jets because synchrotron emission is naturally polarized, with the inferred direction of the magnetic field vector indicating the weighted direction of the magnetic field in the radiating volume. In FR~IIs, where the radio-optical spectrum cannot be neatly extrapolated to the X-rays, high-energy synchrotron emission requires a second electron population. The existence of such a component would drastically alter our picture of FR~II jets, which until recently were not believed to accelerate electrons to $\gamma>10^6$.  This would require highly efficient particle acceleration mechanisms that can operate well outside the host galaxy (e.g., in PKS~1136$-$135 the X-ray emitting knots are at projected distances of $30-60$ kpc from the AGN).  If the optical and X-ray emission is synchrotron radiation, the optical polarization will be high, comparable to that seen in the radio, but with characteristics that may be linked to acceleration processes.

\begin{figure}
  \includegraphics[height=.55\textheight,angle=-90]{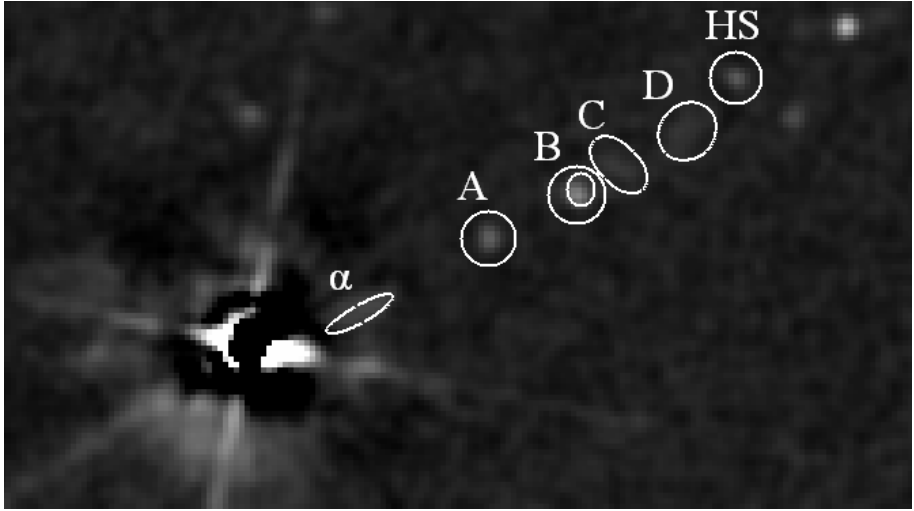}
  \caption{PSF and galaxy subtracted Stokes~I image of the PKS~1136$-$135 jet. Initially selected apertures are shown as white contours. Pixels  with $I<2\sigma_{I}$ are excluded from apertures.}
\end{figure}

The second possibility is inverse-Comptonization of cosmic microwave background photons (IC-CMB), \cite{Celotti:01}.  This requires a jet that remains highly relativistic out to distances of hundreds of kpc \cite{Tavecchio:00,Georg:03}, viewed within a few degrees of the line of sight.  Any optical IC-CMB would be linked to nearly cold electrons, with $\gamma<10$, a population of particles that has never been tracked before.  If the emitter is moving at relativistic bulk speeds, $\Gamma \gg 1$, then the forward-bunching effect will make the CMB photons essentially unidirectional in the jet frame. The IC scattering on the unidirectional and unpolarized photon beam by high-energy electrons having large Lorentz factors ($\gamma \gg 1$) should be unpolarized. On the other hand, the scattered radiation by cold electrons ($\gamma \sim 1$; so-called Bulk Comptonization) in the jet is expected to be highly polarized \cite{Begelman:87}.  We have carried out calculations covering the intermediate regime with $\gamma \sim$~few, making use of the general expression for the intensity and polarization of singly-scattered Comptonized radiation presented by \cite{Poutanen:93}. We found that, for a power-law energy distribution of electrons with a cutoff at $\gamma_{\min} = 2$, the polarization degree can be as large as $8\%$ with the direction of the electric field vector perpendicular to the jet axis (see also \cite{McNamara:09}).

A second Comptonization process is also possible.  This is synchrotron self-Compton (SSC) radiation, in which the seed photons come from the jet's low-frequency radio emission.  While SSC is unavoidable, it is unlikely to dominate the X-ray emission of the jet knots because in order to fit the observed X-ray emission, one requires a jet that is massively out of equipartition (by factors of $20-100$).  However, SSC is the leading scenario for X-ray emission from the terminal hotspots of powerful jets \cite{Harris:94,Wilson:01}. SSC predicts optical polarization properties similar to that of the lowest-frequency radio emission.

\section{Results and Discussion}

Polarization observations have now been done with HST for the jet of PKS~1136$-$135. In Figure~1, we show the Stokes I image from those observations.  In Table 1, we give polarization properties
for the jet components in apertures shown in Figure~1.  We will discuss the details of the observations and data reduction procedures in a future paper.  

Knot~A is shown to be highly polarized, with fractional polarization $\Pi=36 \pm 6\%$, and an inferred magnetic field vector close to that seen in the radio.  Knot~B, however, is weakly polarized, with $\Pi < 11 \%$ at $2 \sigma$ and only in the central region of the knot. Moreover, the magnetic field vector in knot~B is perpendicular to the direction seen in radio.  This indicates that different emission mechanisms dominate in these two regions in the optical as well as X-rays.  Knot~A is clearly dominated by synchrotron emission up to energies of $10\keV$.  This is the first time high-energy synchrotron emission has been proven for any quasar jet at distances of kiloparsecs from the AGN.  Knot~B, however, is consistent with the IC-CMB mechanism dominating. For the first time we also detect  knots~$\alpha$, C, and D in optical. Knot~$\alpha$, like knot~B, is weakly polarized with the direction  of the magnetic field vector perpendicular to the one seen in radio, while knots~C and D are highly polarized ($\Pi > 60\%$) with magnetic field position angle (MFPA) similar to the one in radio in agreement with the synchrotron  radiation being the primarily emission mechanism in these knots in optical.

\begin{table}[htb]
\begin{tabular}{lcccccc}
\hline
  Knot: & $\alpha$ & A & B\tablenote{Inner aperture} & C & D & HS \\
\hline
$\Pi\;(\%)$: & $13\pm 7$ & $36\pm 6$ & $11\pm 5$ & $88\pm 11$ & $57\pm 11$ & $26\pm 8$ \\
$\mbox{MFPA}\;(\arcdeg)$ & $47\pm 15$ & $-49\pm 4$ & $61\pm 10$ & $-24\pm 3$ & $-20\pm 5$ & $-83\pm 8$\\
\hline
\end{tabular}
\caption{The results of HST aperture polarimetry for PKS~1136$-$136.
}
\label{tab:a}
\end{table}

\begin{theacknowledgments}
Jet research at FIT is supported by NASA LTSA grant NNX07AM17G and STScI grant GO-11138.01.
\end{theacknowledgments}

\end{document}